\documentclass{ptptex}

\usepackage{graphicx}

\newcommand{\eq}{\begin{eqnarray}}
\newcommand{\en}{\end{eqnarray}}

\def \mate<#1|#2|#3>{\mbox{$\langle {#1}|\,{#2}\,|{#3}\rangle$}}

\title{Magnetic Moments of S-Shell Pentaquarks\\ 
       in the Constituent Quark Model
}

\author{ T. \textsc{Inoue}, 
         V. E. \textsc{Lyubovitskij},\footnote{On leave of absence from the 
         Department of Physics, Tomsk State University, 634050 Tomsk, Russia.} 
         Th. \textsc{Gutsche} and Amand \textsc{Faessler} 
}

\inst{Institut f\"ur Theoretische Physik, Universit\"at T\"ubingen,\\
         Auf der Morgenstelle 14, D-72076 T\"ubingen, Germany
}

\abst{
We study the magnetic moments of the recently discovered exotic baryons 
$\Theta^+$ and $\Xi^{--}$ and their multiplet partners in the framework 
of a naive additive quark model. These baryons are formulated as pentaquark 
states in which four quarks and a single antiquark exist in their 
ground state orbit. We evaluate the magnetic moments of both 
$J^P=1/2^-$ and $J^P=3/2^-$ multiplets, and elucidate the differences. 
}                           

\begin{document}

\maketitle

Recently, two exotic baryons, $\Theta^+$ and $\Xi^{--}$, have been observed 
in experiments~\cite{Nakano:2003qx}, but there remains some debate 
about their possible existence (see, e.g., Refs.~\citen{Arndt:2003dm}). 
Theoretical interpretations of these baryons have been given in various 
scenarios: In addition to the chiral soliton quark model and the hadron bound 
state approaches, many valence pentaquark models have been proposed,
including a wide variety of correlated and uncorrelated quark configurations. 
For a review, see Ref.~\citen{Jennings:2003wz} and references therein. 
Pentaquark models without stringent dynamical constraints can in general 
describe an enormous number of configurations that obey Fermi statistics
and satisfy the color singlet condition. This, in turn, leads to a huge 
number of pentaquark states arranged in various multiplets. 
However, this is phenomenologically unrealistic. 
For this reason, at the present time, in the construction of models, 
some preferred configurations are selected, and treatment of full spectrum is not carried out.
This situation is in contrast to that regarding the valence 
quark model for conventional baryons.

The valence quark picture for the so-called ground state baryons, 
for which three quarks are in the S-wave ground state and form a symmetric 
representation of spin-flavor $SU(6)$ in accordance with Fermi statistics and 
the color singlet condition, is believed to have a certain validity. 
The most compelling reason, in addition to its successful classification of baryons, 
is that the picture naturally account for the pattern of the magnetic 
moments~\cite{Beg:1964nm}. For example, the naive constituent quark model 
predicts a neutron-proton ratio of $\mu_n/\mu_p = -2/3$, which is rather 
close to the experimental ratio of $\mu_n/\mu_p=-0.68$~\cite{Eidelman:2004wy}. 
The other ratios within the baryon octet are also reproduced quite nicely 
with a reasonable flavor $SU(3)$ breaking. 
(Details are given, for example, in Ref.~\citen{Donoghue:dd}.)
In analogy, a realistic pentaquark model (if it exists) should also be able 
to reproduce the corresponding magnetic moments, as in the case of the 
conventional three-quark baryonic states. Although experimental data concerning 
exotic baryons are lacking at this time, the situation is expected 
to change in the near future~\cite{Zhao:2003gs}. Several studies on the 
magnetic moments of pentaquark states have already been carried out.
A study of the magnetic moments of pentaquark systems using several models,
including correlated and uncorrelated ones, is presented in Ref.~\citen{Liu:2003ab}. 
Magnetic moments have been also predicted in the constituent quark approach in Ref.~\citen{Bijker:2004gr}.
In Refs.~\citen{Kim:2003ay,Kim:2004eh}, the magnetic moments of the antidecuplet baryon multiplet
are worked out in the chiral quark soliton model, whereas a QCD sum rule evaluation for $\Theta^+$
is carried out in Ref.~\citen{Huang:2003bu}. In case that experimental data exist
for the magnetic moments of  $\Theta^+$ and possibly $\Xi^{--}$, 
different predictions serve to further constrain the interpretation of the exotic baryons. 

In this paper, we extend the analysis of the magnetic moments of pentaquark 
systems in the constituent quark approach given in Refs.~\cite{Bijker:2004gr} 
with spin-parity $J^P = 1/2^-, 1/2^+$ and $3/2^+$. 
In particular, we include the members of the antidecuplet with $J^P = 3/2^-$.
Based on the success of this simple model for conventional three-quark baryonic states~\cite{Beg:1964nm},
such considerations appear to be reasonable as a starting point in pentaquark models. 
These tree level results also serve as an input for more detailed 
consideration of pentaquark physics in sophisticated theoretical approaches, 
like, e.g., the perturbative chiral quark model (PCQM)~\cite{Lyubovitskij:2001nm}. 

We start with the standard definition of the magnetic moment of 
a ground state baryon $B$ in the impulse approximation  
(see Refs.~\cite{Beg:1964nm} and \cite{Donoghue:dd} for details): 
\eq 
\mu_{B} \, = \,  \sum\limits_{i=1}^{3} \, \mate<B| \hat{\mu}_i  \, Q_i \, 
\sigma_i^3 |B> \,, 
\label{eqn:barmag}
\en
where $\hat{\mu}_i$, $Q_i$ and $\sigma_i^3$ are the reduced magnetic moment, 
charge and spin of individual quark. 
We fit the values for the reduced quark magnetic moments $\hat{\mu}_i$ 
using the experimental values of the magnetic moments of the proton 
$(\mu_p = 2.793 \, \mu_N)$, neutron $(\mu_p = -1.913 \, \mu_N)$ 
and $\Lambda$-hyperon $(\mu_p = -0.613 \, \mu_N)$ 
and their spin-flavor $SU(6)$ wave functions: 
\eq\label{mag_quarks}
\hat \mu_u = 2.78 \mu_N \,, \ \hat \mu_d = 2.92 \mu_N \,,  \ 
\hat \mu_s = 1.84 \mu_N \,. 
\en
Analogously, in the constituent quark model, the magnetic moments of the pentaquark 
baryons are given by the sum of the single quark and antiquark magnetic 
moments. When all valence particles are in the ground state S-wave orbit, 
the magnetic moment of the pentaquark baryon $B$ is given by 
\eq\label{master_eq} 
\mu_{B} \, = \,  \sum\limits_{i=1}^{5} \, 
\mate<B| \hat{\mu}_i  \, Q_i \, \sigma_i^3 |B>\,.
\en
This expression is exploited in order to obtain a prediction for the 
magnetic moments of the pentaquark baryons depending on the ansatz for the 
color-spin-flavor wave functions and the fit values of $\hat{\mu_i}$. 

\begin{table}[tbp]
\begin{center}
\caption{ Matrix element 
  $\mate<1^C\!,8^F\!,(\frac12,\frac12)^{S}| \,
         \sum\limits_{i=1}^5 \, Q_i \, \sigma^{3}_i \,
        |1^C\!,8^F\!,(\frac12, \frac12)^{S}>$.} 
\label{tbl:tb1}
\begin{tabular}{lrrrrrrr}
\hline \hline
               &  $u$        &    $d$    &    $s$    & $\bar{u}$ & $\bar{d}$ & $\bar{s}$   &  Tot.  \\
\hline
$p_{8}$        & $4/9$       & $-5/{27}$ & $-1/{27}$ &    $0$    & $-2/{27}$ & $-1/{27}$   & $1/9$  \\
$n_{8}$        & ${10}/{27}$ & $-2/9$    & $-1/{27}$ & $4/{27}$  &      $0$  & $-1/{27}$   & $2/9$  \\
$\Lambda_{8}$  & $1/3$       & $-1/6$    & $-1/9$    & $1/9$     & $-1/{18}$ & $0$         & $1/9$  \\
$\Sigma_{8}^+$ & $4/9$       & $-1/{27}$ & $-5/{27}$ &    $0$    & $-1/{27}$ & $-2/{27}$   & $1/9$  \\
$\Sigma_{8}^0$ & $7/{27}$    & $-7/{54}$ & $-5/{27}$ & $1/{27}$  & $-1/{54}$ & $-2/{27}$   & $-1/9$ \\
$\Sigma_{8}^-$ & $2/{27}$    & $-2/9$    & $-5/{27}$ & $2/{27}$  &      $0$  & $-2/{27}$   & $-1/3$ \\
$\Xi_{8}^0$    & ${10}/{27}$ & $-1/{27}$ & $-2/9$    & $4/{27}$  & $-1/{27}$ &    $0$      & $2/9$  \\
$\Xi_{8}^-$    & $2/{27}$    & $-5/{27}$ & $-2/9$    & $2/{27}$  & $-2/{27}$ &    $0$      & $-1/3$ \\
 \hline
\end{tabular}
\end{center}

\begin{center}
\caption{ Matrix element 
  $\mate<1^C\!,8^F\!,(\frac32, \frac32)^{S}| \,
         \sum\limits_{i=1}^5 \, Q_i \, \sigma^{3}_i \,
        |1^C\!,8^F\!,(\frac32, \frac32)^{S}>$. }
\label{tbl:tb2}
\begin{tabular}{lrrrrrrr}
 \hline \hline
               &  $u$        &    $d$    &    $s$    & $\bar{u}$ & $\bar{d}$ & $\bar{s}$ & Tot.      \\
 \hline 
$p_{8}$        & $2/3$       & $-5/{18}$ & $-1/{18}$ &    $0$    & $2/9$     & $1/9$     & $2/3$     \\
$n_{8}$        & $5/9$       & $-1/3$    & $-1/{18}$ & $-4/9$    &   $0$     & $1/9$     & $-1/6$    \\
$\Lambda_{8}$  & $1/2$       & $-1/4$    & $-1/6$    & $-1/3$    & $1/6$     & $0$       & $-1/{12}$ \\
$\Sigma_{8}^+$ & $2/3$       & $-1/{18}$ & $-5/{18}$ &    $0$    & $1/9$     & $2/9$     & $2/3$     \\
$\Sigma_{8}^0$ & $7/{18}$    & $-7/{36}$ & $-5/{18}$ & $-1/9$    & $1/{18}$  & $2/9$     & $1/{12}$  \\
$\Sigma_{8}^-$ & $1/9$       & $-1/3$    & $-5/{18}$ & $-2/9$    &      $0$  & $2/9$     & $-1/2$    \\
$\Xi_{8}^0$    & $5/9$       & $-1/{18}$ & $-1/3$    & $-4/9$    & $1/9$     &    $0$    & $-1/6$    \\
$\Xi_{8}^-$    & $1/9$       & $-5/{18}$ & $-1/3$    & $-2/9$    & $2/9$     &    $0$    & $-1/2$    \\
 \hline
\end{tabular}
\end{center}

\begin{center}
\caption{ Matrix element   
  $\mate<1^C\!,\bar{10}^F\!,(\frac12, \frac12)^{S}| \,
         \sum\limits_{i=1}^5 \, Q_i \, \sigma^{3}_i \,
        |1^C\!,\bar{10}^F\!,(\frac12, \frac12)^{S}>$. }
\label{tbl:tb3}
\begin{tabular}{lrrrrrrr}
\hline \hline
                      &    $u$   &    $d$    &    $s$    &  $\bar{u}$ & $\bar{d}$ & $\bar{s}$ & Tot.  \\
\hline
$\Theta^+$            & $4/9$    & $-2/9$    &    $0$    &    $0$     &      $0$  & $-1/9$    & $1/9$ \\
$p_{\bar{10}}$        & $4/9$    & $-4/{27}$ & $-2/{27}$ &    $0$     & $-1/{27}$ & $-2/{27}$ & $1/9$ \\
$n_{\bar{10}}$        & $8/{27}$ & $-2/9$    & $-2/{27}$ & $2/{27}$   &      $0$  & $-2/{27}$ &   $0$ \\
$\Sigma_{\bar{10}}^+$ & $4/9$    & $-2/{27}$ & $-4/{27}$ &    $0$     & $-2/{27}$ & $-1/{27}$ & $1/9$ \\
$\Sigma_{\bar{10}}^0$ & $8/{27}$ & $-4/{27}$ & $-4/{27}$ & $2/{27}$   & $-1/{27}$ & $-1/{27}$ &   $0$ \\
$\Sigma_{\bar{10}}^-$ & $4/{27}$ & $-2/9$    & $-4/{27}$ & $4/{27}$   &      $0$  & $-1/{27}$ & $-1/9$\\
$\Xi_{\bar{10}}^+$    & $4/9$    &      $0$  & $-2/9$    &    $0$     & $-1/9$    &    $0$    & $1/9$ \\
$\Xi_{\bar{10}}^0$    & $8/{27}$ & $-2/{27}$ & $-2/9$    & $2/{27}$   & $-2/{27}$ &    $0$    &   $0$ \\
$\Xi_{\bar{10}}^-$    & $4/{27}$ & $-4/{27}$ & $-2/9$    & $4/{27}$   & $-1/{27}$ &    $0$    & $-1/9$\\
$\Xi_{\bar{10}}^{--}$ &    $0$   & $-2/9$    & $-2/9$    & $2/9$      &      $0$  &    $0$    & $-2/9$\\
 \hline
\end{tabular}
\end{center}

\begin{center}
\caption{ Matrix element 
  $\mate<1^C\!,\bar{10}^F\!,(\frac32, \frac32)^{S}| \,
         \sum\limits_{i=1}^5 \, Q_i \, \sigma^{3}_i \,
        |1^C\!,\bar{10}^F\!,(\frac32, \frac32)^{S}>$. }
\label{tbl:tb4}
\begin{tabular}{lrrrrrrr}
\hline \hline
                      &  $u$  &    $d$  &    $s$  & $\bar{u}$ & $\bar{d}$ & $\bar{s}$ & Tot.   \\
\hline
$\Theta^+$            & $2/3$ & $-1/3$  &    $0$  &    $0$    &  $0$      & $1/3$     & $2/3$  \\
$p_{\bar{10}}$        & $2/3$ & $-2/9$  & $-1/9$  &    $0$    & $1/9$     & $2/9$     & $2/3$  \\
$n_{\bar{10}}$        & $4/9$ & $-1/3$  & $-1/9$  & $-2/9$    &  $0$      & $2/9$     &   $0$  \\
$\Sigma_{\bar{10}}^+$ & $2/3$ & $-1/9$  & $-2/9$  &    $0$    & $2/9$     & $1/9$     & $2/3$  \\
$\Sigma_{\bar{10}}^0$ & $4/9$ & $-2/9$  & $-2/9$  & $-2/9$    & $1/9$     & $1/9$     &   $0$  \\
$\Sigma_{\bar{10}}^-$ & $2/9$ & $-1/3$  & $-2/9$  & $-4/9$    &  $0$      & $1/9$     & $-2/3$ \\
$\Xi_{\bar{10}}^+$    & $2/3$ &  $0$    & $-1/3$  &    $0$    & $1/3$     &  $0$      & $2/3$  \\
$\Xi_{\bar{10}}^0$    & $4/9$ & $-1/9$  & $-1/3$  & $-2/9$    & $2/9$     &  $0$      &   $0$  \\
$\Xi_{\bar{10}}^-$    & $2/9$ & $-2/9$  & $-1/3$  & $-4/9$    & $1/9$     &  $0$      & $-2/3$ \\
$\Xi_{\bar{10}}^{--}$ &  $0$  & $-1/3$  & $-1/3$  & $-2/3$    &  $0$      &  $0$      & $-4/3$ \\
 \hline
\end{tabular}
\end{center}
\end{table}

In Ref.~\citen{Inoue:2004cc}, we studied a configuration, originally 
studied in Ref.~\citen{Carlson:2003pn}, which was motivated by 
experimental result suggesting that the $\Theta^+$ baryon is an isoscalar~\cite{Nakano:2003qx}. 
Here, the spin and isospin coupling of the subsystem of the four non-strange quarks is chosen to have
total spin $S=1$ and isospin $I=0$. Considering all flavor contributions, 
36 pentaquark states are distributed among four multiplets: two flavor 
$SU(3)$ octets and two antidecuplets both with the two possible spin-parities $J^P = 1/2^-$ and $3/2^-$. 
(For the details of this configuration, see Refs.~\citen{Inoue:2004cc,Carlson:2003pn}.)
We remark here that the first calculation of the wave functions of the S-shell pentaquark states 
with $J^P = 1/2^-, 3/2^-$ was done in Ref.~\citen{Strottman:1979qu} in 
the context of MIT bag model. Our color-spin-flavor parts for these states 
coincide with the results of Ref.~\citen{Strottman:1979qu}. 
It is also worth noting that the subsystem of the S-shell $\Theta^+$ pentaquark
can be coupled also in the $I=1$ and $I=2$ states, where $\Theta^+$ belongs to flavor 
27-plet and 35-plet, respectively. These multiplets are not studied in this paper.
(Further discussion of them can be found in Refs.~\citen{Bijker:2003pm,Oh:2004gz,Capstick:2003iq}.)

We first consider the flavor symmetry limit and the quark-antiquark 
symmetry limit, in which, the reduced magnetic moments of the quark and antiquark 
satisfy the relation $\hat\mu_q = \hat \mu_{\bar q}$. In this simplest case,
the magnetic moment of a pentaquark is proportional to the matrix 
element $\mate<B| \sum_{i=1}^{5} \, Q_i \, \sigma_i^3 |B>$. 
In Tables \ref{tbl:tb1}-\ref{tbl:tb4} we list these matrix elements for the given configurations. 
Note that our Table \ref{tbl:tb3} is identical to Table 1 ($J^P=1/2^-$ part) of Ref.~\citen{Bijker:2004gr}.

We start with the $J^P=1/2^-$ antidecuplet using Table \ref{tbl:tb3}. 
All matrix elements are proportional to the baryon charge $Q_{B}$, with 
\eq 
 \mate<B(\bar{10},1/2^-)\!\uparrow|\sum\limits_{i=1}^5 Q_i 
\sigma^{3}_i|B(\bar{10},1/2^-)\uparrow> = 1/9 \, {Q_{B}}\,.
\en
Hence, the magnetic moments also follow this. 
This result was originally obtained in Ref.~\citen{Bijker:2004gr}.
In Ref.~\citen{Liu:2003ab} it is shown that the magnetic moments of the 
antidecuplet pentaquarks following the Jaffe-Wilczek configuration~\cite{Jaffe:2003sg} 
are also proportional to the baryon charge in the flavor symmetry limit. 
Moreover, the same feature is found in the chiral quark soliton model~\cite{Kim:2003ay}. 
The charge proportionality is a common aspect of the flavor $SU(3)$ spin $1/2$ antidecuplet
with baryon number one, and also for the spin 3/2 antidecuplet, as evident from Table \ref{tbl:tb4}. 
This, however, make it more difficult to distinguish between models when 
studying magnetic moments. 
For the $\Theta^+(1/2^-)$ state, we have 
\eq 
\mate<\Theta^+(1/2^-)\!\uparrow|\sum\limits_{i=1}^5 Q_i 
\sigma^{3}_i|\Theta^+(1/2^-)\!\uparrow>=1/9\,, 
\en
where the corresponding matrix element for the proton is unity. 
Thus the magnetic moment of $\Theta^+(1/2^-)$ is obtained as $\mu_p/9$,
giving a small positive value of about $0.3 \, \mu_N$ in the symmetry limit. 
As pointed out in Ref.~\citen{Bijker:2004gr}, this prediction of the simple quark model is 
consistent with that of the chiral quark soliton model~\cite{Kim:2003ay} and 
a QCD sum rule calculation~\cite{Huang:2003bu}. 
However, the authors of Ref.~\citen{Kim:2003ay} recently revised their result with a new fit, 
according to which, the magnetic moment of the $\Theta^+$ baryon becomes small and negative~\cite{Kim:2004eh}. 
For $\Xi^{--}$ baryon, magnetic moment is predicted to be $-0.6 \, \mu_N$ 
in the present calculation.

Next, we discuss the $J^P=3/2^-$ antidecuplet, based on Table \ref{tbl:tb4}.
The corresponding magnetic moments are again proportional to the baryon 
charge. The  matrix elements are 6 times larger than those for the spin 1/2 states. 
Now, the magnetic moment of the $\Theta^+(3/2^-)$ states is $1.8 \, \mu_N$ 
in the flavor symmetry limit. The spin 3/2 states always accompany the spin 
1/2 states in pentaquark models. In Ref.~\citen{Inoue:2004cc} the spin 3/2 
states are approximately 170 MeV heavier than the spin 1/2 states due to the 
semi-perturbative gluon effect. Assigning the lighter spin 1/2 state to the 
observed $\Theta^+$ baryon, one would also expects a $3/2^-$ $\Theta^+$ at about 
1710 MeV. However, in general, a spin of 3/2 cannot be ruled out for 
the observed $\Theta^+$ baryon.  Moreover, this value of the spin is also
favored by phenomenological consideration, since a $3/2^-$ $\Theta^+$ couple to the  
D-wave $KN$ channel, leading to a small width. 
Note that the $\Lambda(1520)$ with $J^P = 3/2^-$ decays mainly into
the D-wave $\bar K N$ and $\pi\Sigma$, and the total width is about 15 MeV~\cite{Eidelman:2004wy}. 
The present model predicts a large magnetic moment of $\Xi_{\bar{10}}^{--}(3/2^-)$,
(about $-3.6 \, \mu_N$), which, in turn, enhances photo-production. 

The matrix elements corresponding to the $J^P=1/2^-$ and $3/2^-$ flavor octet
pentaquark baryons are given in Tables \ref{tbl:tb1} and \ref{tbl:tb2}. 
In this case, the magnetic moments are not proportional to the baryon charge. 
The results for two octets differ qualitatively from each other
and also from the result for the conventional three-quark octet. 
The set of magnetic moments is therefore characteristic for each octet. 
The set of magnetic moments in each multiplet satisfies a simple sum rule, 
namely, that the sum of the magnetic moments equals zero. 
This sum rule seems to be universal for baryon multiplets including canonical and exotic ones. 
The set of moments of the Jaffe-Wilczek octet pentaquarks 
also satisfy the sum rule in the flavor symmetry limit~\cite{Liu:2003ab}.
Moreover, in the limit that the masses of the quark and diquark become identical,
the moments coincide with the present results for the spin 1/2 octet. 

\begin{figure}[tbp]
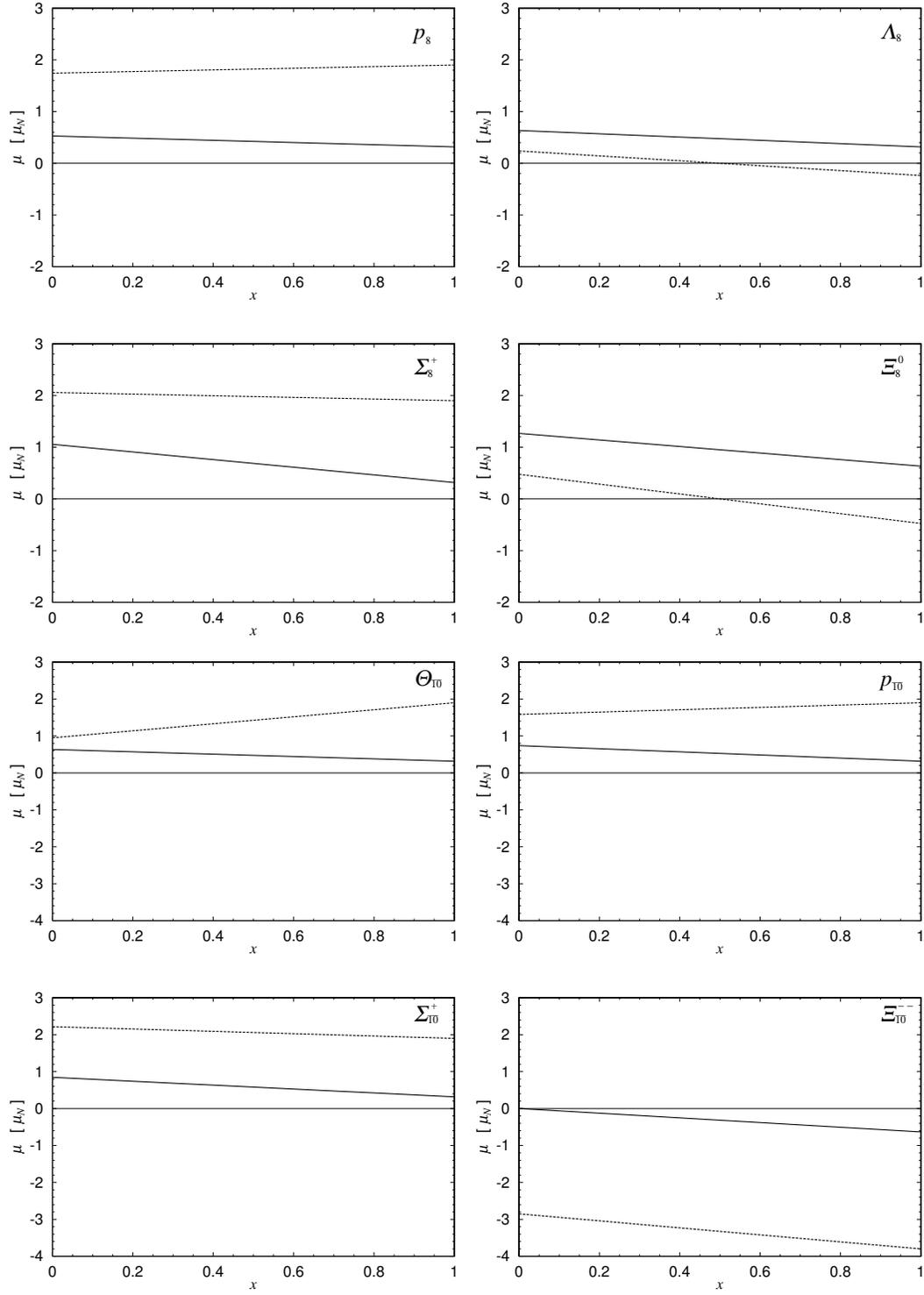

 \begin{center}
    \includegraphics[width=138mm]{fig1.epsi}

\vspace*{.3cm}

    \includegraphics[width=138mm]{fig2.epsi}
  \caption{
    Magnetic moments of the octet and antidecuplet pentaquark baryons 
    in units of $\mu_N$ as functions of the $SU(3)$ breaking  parameter $x=\hat{\mu}_s/\hat{\mu}$. 
    The solid line represents $J^P=1/2^-$, and the dashed line represents $J^P=3/2^-$.}
  \label{fig:mag}
 \end{center}
\end{figure}

Next, we study the effect of flavor $SU(3)$ symmetry breaking on the 
predictions. While we retain the isospin symmetry, the breaking
is taken into account with $\hat{\mu} = \hat{\mu}_u = \hat{\mu}_d \ne \hat{\mu}_s$. 
Some of the results are plotted as functions of the ratio $x = \hat{\mu}_s / \hat{\mu}$  
in Fig.~\ref{fig:mag} for the octet and the antidecuplet. 
The reduced magnetic moments of the non-strange quarks are fixed as
$\hat{\mu} = (2.78 + 2.92)/2 \cdot \mu_N = 2.85 \mu_N$. 
The value $x=1$ corresponds to the flavor symmetry limit, 
while the value $x=0$ corresponds to $\hat\mu_s = 0$. 
Judging from the data for ground state baryons, 
a value of $x \simeq 0.65$ is favored [see Eq.~(\ref{mag_quarks})]. 
Although the effect of flavor symmetry breaking is in general not so 
drastic, there can be sizable changes for some magnetic moments. 
For example, for $x = 0.65$, we obtain 
\eq 
& &\mu_{\Theta^+(1/2^-)} =    0.43 \, \mu_N\,, \hspace*{.5cm}  
   \mu_{\Xi^{--}(1/2^-)} =  - 0.41 \, \mu_N\,, \nonumber \\ 
& &\mu_{\Theta^+(3/2^-)} =    1.57 \, \mu_N\,, \hspace*{.5cm} 
   \mu_{\Xi^{--}(3/2^-)} =  - 3.47 \, \mu_N\,. 
\en
The magnetic moments of some other baryons can be read off 
of Fig.~\ref{fig:mag}. 

The breaking of the flavor symmetry can also induce mixing between 
the flavor multiplets. 
It is therefore expected that the presently considered octet and antidecuplet baryons will mix. 
Because isospin symmetry remains valid, mixing between the octet $\Xi_8^{0}$($\Xi_8^{-}$) 
and the antidecuplet $\Xi_{\bar{10}}^{0}$($\Xi_{\bar{10}}^{-}$) must be negligible.
It is implied, then, that we have two unknown possible mixings 
between $N_{8}$ and $N_{\bar{10}}$ and between $\Sigma_{8}$ and $\Sigma_{\bar{10}}$. 
In Ref.~\citen{Jaffe:2003sg} ideal mixing is assumed for the corresponding baryons.  
We evaluate also the magnetic moments in this limit.
We refer to those with the maximal $s + \bar s$ contents as ``heavy''
and those with minimal $s + \bar s$ contents as ``light''.
The pair $(N_{\mbox{\tiny light}}, N_{\mbox{\tiny heavy}})$ 
is obtained through a $35^{\circ}$ rotation from $(N_{8}, N_{\bar{10}})$,
and the pair $(\Sigma_{\mbox{\tiny light}}, \Sigma_{\mbox{\tiny heavy}})$ 
is obtained through a $55^{\circ}$ rotation from $(\Sigma_{8}, \Sigma_{\bar{10}})$. 
For example, for $x = 0.65$, we obtain 
\eq 
& &\mu_{n_{\mbox{\tiny \,light}}(1/2^-)}      =    1.27 \, \mu_N\,, 
\hspace*{.5cm}  
   \mu_{n_{\mbox{\tiny \,heavy}}(1/2^-)}      = -\,0.41 \, \mu_N\,, 
   \nonumber \\
& &\mu_{\Sigma^+_{\mbox{\tiny light}}(1/2^-)} =    0.43 \, \mu_N\,, 
\hspace*{.5cm}
   \mu_{\Sigma^+_{\mbox{\tiny heavy}}(1/2^-)} =    0.65 \, \mu_N\,. 
\en
The predictions for $\Lambda_{8}$, $\Xi_{8}$, $\Theta_{\bar{10}}$ 
and $\Xi_{\bar{10}}$ are not affected by mixing. 
Note that $\Lambda_8(1/2^-)$ has the same magnetic moment as the 
$\Theta_{\bar{10}}^+(1/2^-)$ for all values of $x$. 

In this paper, we have used the constituent quark approach for simplicity 
because it is, at least, successful in the description of magnetic moments 
of the ground state baryons~\cite{Beg:1964nm,Donoghue:dd}.  
However, in the chiral quark picture, in which valence quarks are supplemented by
a cloud of pseudoscalar mesons, mesonic corrections can be sizable. 
In Refs.~\cite{Lyubovitskij:2001nm}, we developed the perturbative chiral 
quark model on the basis of such a picture, where, for example, in the proton case, 
the valence core contributes about 70\%, while the reminder is accounted for the meson cloud. 
An estimate of meson cloud effects for the $\Theta^+$ baryon gives $\simeq 0.1 \mu_N$,
which is about of 25\% of the valence quark contribution. 
Hence, we expect that the five-quark core gives the leading contribution
to the magnetic moments of pentaquarks, but the chiral corrections are not negligible
(as in the case of conventional three-quark baryons). In a further work, 
we hope to calculate the meson cloud contributions to the magnetic moments 
of all pentaquark states. 

We have extended the study of the magnetic moments of pentaquark states given in Ref.~\citen{Bijker:2004gr}, 
including the multiplet with spin-parity $3/2^-$. 
We applied the constituent quark approach with the parameters fixed by the magnetic moments
of conventional baryons (nucleons and the $\Lambda$-hyperon). 
For the $J^P ={1/2}^-$ antidecuplet, we obtain  $0.43 \, \mu_N$ for 
the $\Theta^+(1/2^-)$ state and $- 0.41 \, \mu_N$ for the $\Xi^{--}({1/2}^-)$ state, 
which may correspond to the observed ones. 
Note that the magnetic moment of $\Theta^+$ is small and positive, 
while for the $\Xi^{--}$ it is small and negative, in the present approach.
The obtained magnetic moment of $\Theta^+({3/2}^-)$ is $1.57 \, \mu_N$,
and that of $\Xi^{--}({3/2}^-)$ is $- 3.46 \, \mu_N$. 
These states are expected to accompany the possibly observed $1/2^-$ states. 
In the flavor symmetry limit, the magnetic moments of the antidecuplet states are 
proportional to their charges, and the magnetic moments of a multiplet sums to zero. 
We have studied also the octet of pentaquark baryons. 
We found that each octet has a characteristic set of magnetic moments
and also satisfies the sum rule in the flavor symmetry limit. 
Although the pentaquark configurations in the present study and in the Jaffe-Wilczek correlated 
model~\cite{Jaffe:2003sg} are completely different, 
the predictions for the magnetic moments obtained from the two models are quite similar. 
Moreover, these predictions coincide with each other when the masses of the quark and the diquark
is equal in the Jaffe-Wilczek model.

\section*{Acknowledgements}
This work was supported by the DFG (contracts FA67/25-3 and GRK683) and  
a President Grant of Russia ``Scientific Schools" (No.~1743.2003). 
This research is also part of the EU Integrated Infrastructure Initiative 
Had\-ron\-phy\-si\-cs Project, under contract number RII3-CT-2004-506078.

\end{document}